%% file: 0PaperID3379.tex
\documentclass[sigconf]{acmart}
\usepackage{booktabs} % For formal tables
\usepackage{amsmath}
\usepackage{amsfonts}
\usepackage{comment}
\usepackage{footmisc}
\usepackage{mathrsfs}
\usepackage{graphicx}
\usepackage{subfigure}
\usepackage{multirow}
\usepackage{algorithm}
\usepackage{algorithmic}
\usepackage{multicol}  
\usepackage{enumitem}
\usepackage{array}
\usepackage{float}
\usepackage{hyperref}

\AtBeginDocument{%
	\providecommand\BibTeX{{%
			\normalfont B\kern-0.5em{\scshape i\kern-0.25em b}\kern-0.8em\TeX}}}

\copyrightyear{2021}
\acmYear{2021}
\setcopyright{acmcopyright}
\acmConference[KDD '21]{Proceedings of the 27th ACM SIGKDD Conference on Knowledge Discovery and Data Mining}{August 14--18, 2021}{Virtual Event, Singapore}
\acmBooktitle{Proceedings of the 27th ACM SIGKDD Conference on Knowledge Discovery and Data Mining (KDD '21), August 14--18, 2021, Virtual Event, Singapore} \acmPrice{15.00}
\acmDOI{10.1145/3447548.3467208}
\acmISBN{978-1-4503-8332-5/21/08}
\begin{document}
\fancyhead{}
\title{AutoLoss: Automated Loss Function Search in Recommendations}

\author{Xiangyu Zhao$^{1,2}$, Haochen Liu$^2$, Wenqi Fan$^{3*}$, Hui Liu$^2$, Jiliang Tang$^2$, Chong Wang$^4$}
\thanks{* Corresponding author.}
\affiliation{
	\institution{$^1$City University of Hong Kong, $^2$Michigan State University, $^3$The Hong Kong Polytechnic University, $^4$Bytedance}
	\country{}
}
\email{{zhaoxi35,liuhaoc1,liuhui7,tangjili}@msu.edu, wenqifan@polyu.edu.hk, chong.wang@bytedance.com}
%\email{{wguo,jshi,sidwang,hgao}@linkedin.com, bo.long@gmail.com}

\renewcommand{\shortauthors}{Xiangyu Zhao, et al.}

\begin{abstract}
Designing an effective loss function plays a crucial role in training deep recommender systems. Most existing works often leverage a predefined and fixed loss function that could lead to suboptimal recommendation quality and training efficiency. Some recent efforts rely on exhaustively or manually searched weights to fuse a group of candidate loss functions, which is exceptionally costly in computation and time. They also neglect the various convergence behaviors of different data examples. In this work, we propose an AutoLoss framework that can automatically and adaptively search for the appropriate loss function from a set of candidates. To be specific, we develop a novel controller network, which can dynamically adjust the loss probabilities in a differentiable manner. Unlike existing algorithms, the proposed controller can adaptively generate the loss probabilities for different data examples according to their varied convergence behaviors. Such design improves the model's generalizability and transferability between deep recommender systems and datasets. We evaluate the proposed framework on two benchmark datasets. The results show that AutoLoss outperforms representative baselines. Further experiments have been conducted to deepen our understandings of AutoLoss, including its transferability, components and training efficiency.
\end{abstract}




\maketitle
\input{1Introduction}
\input{2Framework}
\input{3Experiments}
\input{5RelatedWork}
\input{6Conclusion}
\vspace{-2.1mm}
\section*{ACKNOWLEDGEMENTS}
This work is supported by National Science Foundation (NSF) under grant numbers IIS1907704, IIS1928278, IIS1714741, IIS1715940, IIS1845081, CNS1815636, and an internal research fund from the Hong Kong Polytechnic University (project no. P0036200). 
\bibliographystyle{ACM-Reference-Format}
\bibliography{9Reference}
\end{document}

%% file: 1Introduction.tex
\section{Introduction}
\label{sec:introduction}
In the era of information explosion, recommender systems play a pivotal role in alleviating information overload, which vastly enhance user experiences in many commercial applications, such as generating playlists in video and music services~\cite{zhao2021dear,zhao2020jointly}, recommending products in online stores~\cite{zhao2020whole,zou2020neural,fan2020attacking,zhao2018deep,zhao2018recommendations}, and suggesting locations for geo-social events~\cite{liu2017experimental,zhao2016exploring,guo2016cosolorec}. With the recent growth of deep learning techniques, there have been increasing interests in developing deep recommender systems (DRS)~\cite{nguyen2017personalized,wu2016personal}. DRS has improved the recommendation quality since they can effectively learn feature representations and capture the nonlinear relationships between users and items via deep architectures~\cite{zhang2019deep}. Aside from developing sophisticated neural network architectures, well-designed loss functions have also been demonstrated to be effective in improving the performance in different recommendation tasks, such as item rating prediction (regression)~\cite{ravi2016collaborative}, click-through rate prediction (binary classification)~\cite{guo2017deepfm,ge2021towards}, user behavior prediction (multi-class classification)~\cite{zhao2019toward}, and item retrieval (clustering)~\cite{gao2020deep}. 

To optimize DRS frameworks, most existing works are based on a predefined and fixed loss function, such as \textit{mean-squared-error} (MSE) or \textit{mean-absolute-error} (MAE) loss for regression tasks. Then DRS frameworks are optimized in a back-propagation manner, which computes gradients effectively and efficiently to minimize the given loss on the training dataset. During this process, the key step is to calculate gradients of network parameters for minimizing loss functions. However, it is often unclear whether the gradients generated from a given loss function are optimal. For example, in regression tasks, the MSE loss can ensure that the trained model has no outlier predictions with huge errors, while MAE performs better if we only want a well-rounded model that performs well on the majority~\cite{fahrmeir2007regression,chatterjee2015regression}.
Therefore, solely utilizing a predefined and fixed loss function for all \textit{data examples}, i.e., \textit{user-item interactions}, cannot guarantee the optimal gradients throughout, especially when the interactions have varied convergence behaviors in the non-stationary environment of online recommendation platforms. 
In addition, there is often a gap between the model training and evaluation performance in real-world recommender systems. For instance, we usually train a predictive model by minimizing \textit{cross-entropy loss} in online advertising, and evaluate the model performance by \textit{click-through rate} (CTR). Consequently, it naturally raises a question - can we incorporate more loss functions in the training phase to enhance the model performance?

\begin{figure*}
	\centering
	%	\hspace*{-7mm}1.049
	\includegraphics[width=0.92\linewidth]{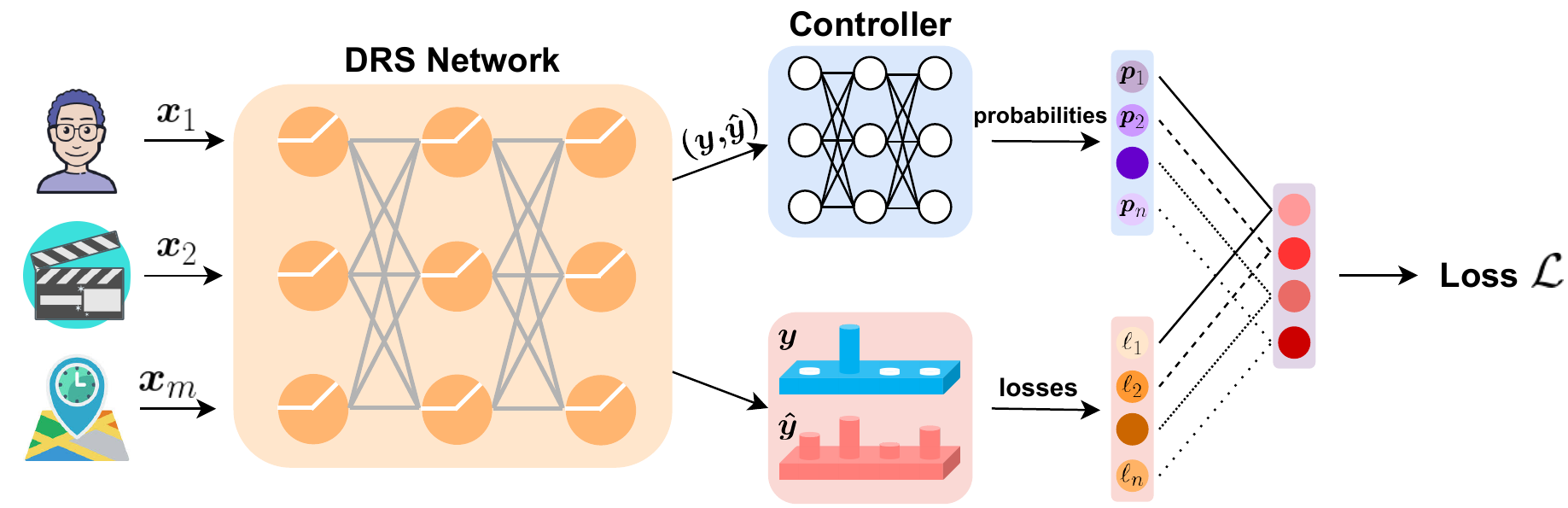}
	\caption{Overview of the AutoLoss framework.}
	\label{fig:Fig1_Overview}
	\vspace*{-3mm}
\end{figure*}

Efforts have been made to develop strategies to fuse multiple loss functions, which can take advantage of multiple loss functions in a weighted sum fashion. For example, Panoptic FPN~\cite{kirillov2019panoptic} leverages a grid search to find better loss weights; and UPSNet~\cite{xiong2019upsnet} carefully investigates the weighting scheme of loss functions. However, these works rely on exhaustively or manually search for loss weights from a large candidate space, which would be an extremely costly execution in both computing power and time. Also, they aim to learn a set of unified and static weights over the loss functions, which entirely overlook the different convergence behaviors of data examples. 
Finally, retraining loss weights is always desired when switching among different DRS frameworks or recommendation datasets, which reduces their generalizability and transferability.

In order to obtain more accurate gradients to improve the recommendation performance and the training efficiency, we propose an automated loss function search framework, \textbf{AutoLoss}, which can dynamically and adaptively select appropriate loss functions for training DRS frameworks. Different from existing searching models with predefined and fixed loss functions, or the loss weights exhaustively or manually searched, the optimal loss function in AutoLoss is automatically selected for each data example in a differentiable manner. The experiments on two datasets demonstrate the effectiveness of the proposed framework. We summarize our major contributions as follows: 

\begin{itemize}[leftmargin=*]
	\item We propose an end-to-end framework, AutoLoss, which can automatically select the proper loss functions for training DRS frameworks with better recommendation performance and training efficiency;
%	\vspace{1.6mm}
	\item A novel controller network is developed to adaptively adjust the probabilities over multiple loss functions according to different data examples' dynamic convergence behaviors during training, which enhances the model generalizability between different DRS frameworks and datasets;
%	\vspace{1.6mm}
	\item We empirically demonstrate the effectiveness of the proposed framework on real-world benchmark datasets. Extensive studies verify the importance of model components and the transferability of AutoLoss.
\end{itemize}

The rest of this paper is organized as follows. 
In Section 2, we detail the framework of automatically searching the probabilities over multiple loss functions, the architecture of the main DRS network and controller network, and propose an AutoML-based optimization algorithm. Section 3 carries out experiments based on real-world datasets and presents experimental results. Section 4 briefly reviews related work. Finally, Section 5 concludes this work and discusses future work.

%% file: 2Framework.tex
\section{The Proposed Framework}
\label{sec:framework}
In this section, we will present an end-to-end framework, AutoLoss, which effectively tackles the aforementioned challenges in Section~\ref{sec:introduction} via automatically and adaptively searching the optimal loss function from several candidates according to data examples' convergence behaviors. We will first provide an overview of the framework; next detail the architectures of the main DRS network; then introduce the loss function search method with a novel controller network; and finally provide an AutoML-based optimization algorithm.

\subsection{An Overview}
In this subsection, we will give an overview of the AutoLoss framework. AutoLoss aims to automatically select appropriate loss functions from a set of candidates for different data examples (i.e., user-item interactions). We demonstrate the framework in Figure~\ref{fig:Fig1_Overview}. With a DRS network, a controller network and a set of pre-defined candidate loss functions, the learning process of AutoLoss mainly consists of two major steps. 

{\it The forward-propagation step}. Given a mini-batch of data examples, the main DRS network first generates predictions $\boldsymbol{\hat{y}}$ based on the input features $\boldsymbol{x}$. Then, we can calculate the losses $\{\ell_{i}\}$ for each candidate loss function according to the ground truth labels $\boldsymbol{y}$ and predictions $\boldsymbol{\hat{y}}$. Meanwhile, the controller network takes $\boldsymbol{(y, \hat{y})}$ and outputs the probabilities $\boldsymbol{p}$ over loss functions for each data example. Finally, the overall loss $\mathcal{L}$ can be calculated according to the losses from $\{\ell_{i}\}$ and the probabilities $\boldsymbol{p}$.

{\it The backward-propagation step}. We first fix the parameters of the controller network and update the main DRS network parameters upon the training data examples. Then, we fix the DRS parameters and optimize the controller network parameters based on a mini-batch of validation data examples. This alternative updating approach enhances the generalizability, and prevents  AutoLoss from selecting probabilities that overfit the training data examples~\cite{pham2018efficient,liu2018darts}. Next, we will introduce the details of AutoLoss.

\begin{figure*}
	\centering
	%	\hspace*{-7mm}1.049
	\includegraphics[width=0.76\linewidth]{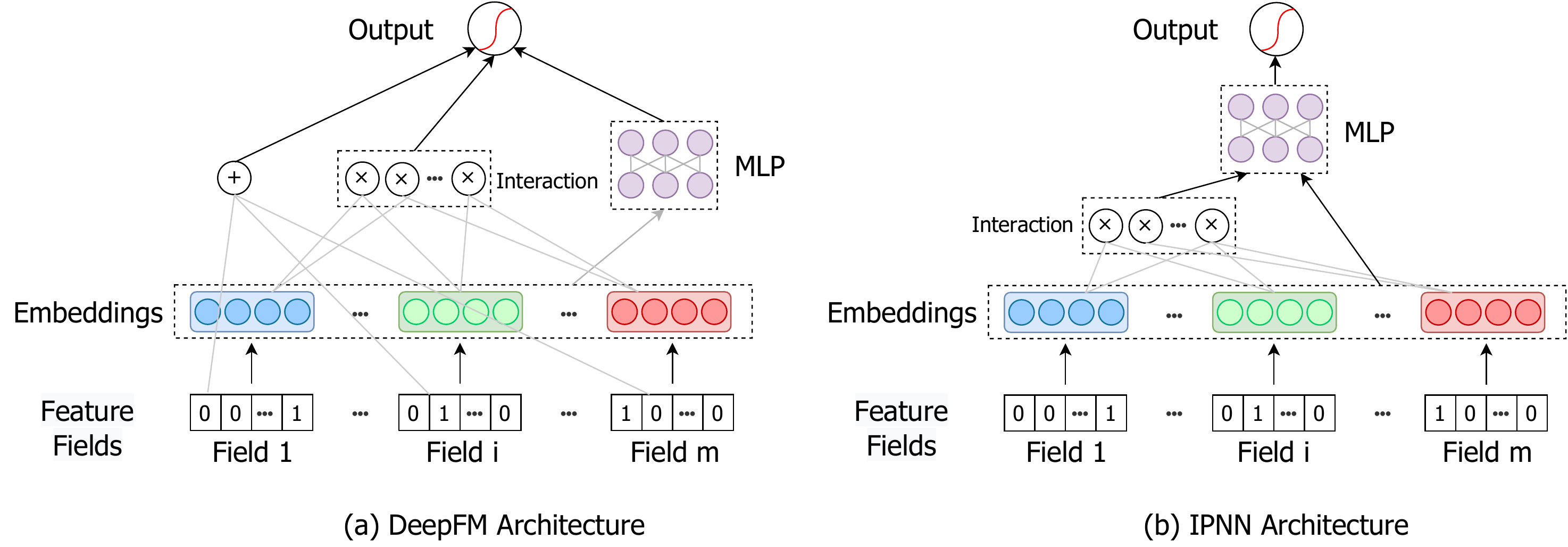}
	\caption{Architectures of DeepFM and IPNN.}
	\label{fig:Fig2_DRS}
	\vspace*{-3mm}
\end{figure*}

\subsection{Deep Recommender System Network}
AutoLoss is quite general for most existing deep recommender system frameworks~\cite{rendle2010factorization,guo2017deepfm,lian2018xdeepfm,qu2016product}. As visualized in Figure~\ref{fig:Fig2_DRS}, they typically have four components: embedding layer, interaction layer, MLP layer and output layer. We now briefly introduce these components.

\subsubsection{\textbf{Embedding Layer}}
The raw input features of users and items are usually categorical or numeric, and in the form of multiple fields. Most DRS works first transform the input features into binary vectors, and then embed them into continuous vectors using a field-wise embedding. In this way, a user-item interaction data example $\boldsymbol{x}=\left[\boldsymbol{x}_{1}, \boldsymbol{x}_{2}, \cdots, \boldsymbol{x}_{m}\right]$ can be represented as the concatenation of binary vectors from all feature fields:
\begin{equation*}
[\underbrace{0,1,0,0, \ldots, 0}_{\boldsymbol{x}_{1}:\,\text {userid }}][\underbrace{1,0}_{\boldsymbol{x}_{2}:\,\text {gender }}][\underbrace{0,1,0,0}_{\boldsymbol{x}_{3}:\,\text {age }}] \overbrace{\cdots \cdots}^{\text {other fields}} \underbrace{[0,1,0,1, \ldots, 0]}_{\boldsymbol{x}_{m}:\,\text {itemid}}
\end{equation*}
where $m$ is the number of feature fields and $\boldsymbol{x}_i$ is the binary vector of the $i^{th}$ field. The categorical data are transformed into binary vectors via one-hot encoding, e.g., $[0,1]$ for ${gender=Female}$ and $[1,0]$ for ${gender=Male}$. The numeric data are first partitioned into buckets, and then we have a binary vector for each bucket, e.g., we can use $[0,0,0,1]$ for child whose $age{\in}[0{,}14]$, $[0,0,1,0]$ for youth whose $age{\in}[15{,}24]$, $[0,1,0,0] $ for adult whose $age{\in}[25{,}64]$, and $[1, 0, 0, 0]$ for seniors whose $age{\geq}65$.

Since vector $\boldsymbol{x}$ is high-dimensional and very sparse, and different feature fields have various lengths, DRS models usually introduce an embedding layer to transform each binary vector $\boldsymbol{x}_i$ into a low-dimensional continuous vector as:
\begin{equation}
\boldsymbol{e}_{i}=\boldsymbol{v}_{i} \boldsymbol{x}_{i}
\end{equation}
where $\boldsymbol{v}_{i} \in R^{d \times u_{i}}$ is the weight matrix with $u_{i}$ the number of unique feature values in the $i^{th}$ feature field, and $d$ is the pre-defined size of low-dimensional vectors\footnote{For multi-valued features (e.g.,``Interest=Movie, Sports''), the feature embedding is the sum or average of multiple embeddings~\cite{covington2016deep}.}. Finally, the embedding layer will output the concatenation of embedding vectors from all feature fields:
\begin{equation}
\boldsymbol{E}=\left[\boldsymbol{e}_{1}, \boldsymbol{e}_{2}, \ldots, \boldsymbol{e}_{m}\right]
\end{equation}

\subsubsection{\textbf{Interaction Layer}}
After representing the input features as low-dimensional embeddings, 
DRS models usually develop an interaction layer to explicitly capture the interactions among feature fields. The most widely used method is factorization machine (FM)~\cite{rendle2010factorization}. In addition to the linear interactions among features, FM can explicitly model the pairwise (second-order) feature interactions via the inner product of feature embeddings:
\begin{equation}
\left[\left\langle \boldsymbol{e}_{1}, \boldsymbol{e}_{2}\right\rangle,\left\langle \boldsymbol{e}_{1}, \boldsymbol{e}_{3}\right\rangle, \ldots,\left\langle \boldsymbol{e}_{m-1}, \boldsymbol{e}_{m}\right\rangle\right]
\end{equation}
where $\langle\cdot, \cdot\rangle$ is the inner product of two embeddings, and the number of pairwise feature interactions is $\boldsymbol{C}_m^2$. Then, the interaction layer will output:
\begin{equation}
l_{fm}=\langle\boldsymbol{w}, \boldsymbol{x}\rangle+\sum_{i=1}^{m} \sum_{j>i}^{m}\left\langle\boldsymbol{e}_{i}, \boldsymbol{e}_{j}\right\rangle
\end{equation}
Where $\boldsymbol{w}$ is the weight over the binary vector $\boldsymbol{x}$ of input features. The first term represents the impact of first-order feature interactions, and the second term reflects the impact of second-order feature interactions. FM can explicitly model even higher order interactions, such as $\sum_{i=1}^{m} \sum_{j>i}^{m} \sum_{t>j}^{m}\left\langle\boldsymbol{e}_{i}, \boldsymbol{e}_{j}, \boldsymbol{e}_{t}\right\rangle$ for third-order, but this will add a lot of computation.

\subsubsection{\textbf{MLP Layer}}
MLP Layer combines and transforms the features, e.g., $\boldsymbol{E}$ and $l_{fm}$, with several fully-connected layers and activations. The output of each layer is:
\begin{equation}
\label{eq:mlp}
\boldsymbol{h}_{l+1}=\operatorname{relu}\left(\boldsymbol{W}_{l} \boldsymbol{h}_{l}+\boldsymbol{b}_{l}\right)
\end{equation}
where $\boldsymbol{W}_{l}$ is the weight matrix and $\boldsymbol{b}_{l}$ is the bias vector for the $l^{th}$ hidden layer. $\boldsymbol{h}_{0}$ is the input of first fully-connected layer, and we denote the final output of MLP layer as $\operatorname{MLP}(\boldsymbol{h}_0)$.

\subsubsection{\textbf{Output Layer}}
Finally, the output layer, which is subsequent to the previous layers, will generate the prediction $\boldsymbol{\hat{y}}$ of a user-item interaction data example. The input $\boldsymbol{h}_{out}$ of output layer can be different in different DRS models, e.g., $\boldsymbol{h}_{out}=[l_{f m}+\mathrm{MLP}(\boldsymbol{E})]$ in DeepFM~\cite{guo2017deepfm} and $\boldsymbol{h}_{out}=\mathrm{MLP}(l_{f m},\boldsymbol{E})$ in IPNN~\cite{qu2016product}, shown in Figure~\ref{fig:Fig2_DRS}. The output layer will yield the prediction $\boldsymbol{\hat{y}}$ of the user-item interaction as:
\begin{equation}
{\boldsymbol{\hat{y}}=\sigma\left(\boldsymbol{W}_{o} \boldsymbol{h}_{out}+\boldsymbol{b}_{o}\right)}
\end{equation}
\noindent where $\boldsymbol{W}_{o}$ and $\boldsymbol{b}_{o}$ are the  weight matrix and bias vector for the output layer. Activation function $\sigma(\cdot)$ is selected based on different recommendation tasks, such as \textit{sigmoid} for binary classification~\cite{guo2017deepfm}, and \textit{softmax} for multi-class classification~\cite{tan2016improved}. Finally, given a set of candidate loss functions, such as mean-squared-error, categorical hinge and cross-entropy, we can compute the candidate losses $\mathcal{L_C}$:
\begin{equation}
\mathcal{L_C} = [\ell_{1}(\boldsymbol{y}, \boldsymbol{\hat{y}}), \ell_{2}(\boldsymbol{y}, \boldsymbol{\hat{y}}), \cdots, \ell_{n}(\boldsymbol{y}, \boldsymbol{\hat{y}})]
\end{equation}
where $\boldsymbol{y}$ is the ground truth label and $n$ is the number of candidate loss functions.

\subsection{Loss Function Search}
AutoLoss aims to adaptively and automatically search the optimal loss function, which can enhance the prediction quality and training efficiency of the DRS network. This is naturally challenging because of the complex relationship between the DRS parameters and the probabilities over candidate loss functions. To address this challenge, many existing works have focused on developing the fusing strategies for multiple loss functions, which can take advantage of multiple loss functions in a weighted sum manner: 
\begin{equation}
\label{eq:overall}
\begin{aligned}
&\mathcal{L}(\boldsymbol{y}, \boldsymbol{\hat{y}}; \boldsymbol{\alpha})=\sum_{i=1}^{n} \alpha_{i}\cdot \ell_{i}(\boldsymbol{y}, \boldsymbol{\hat{y}})\\
\text { s.t. } &\sum_{i=1}^{n} \alpha_{i} =1, \quad \alpha_{i} > 0 \; \forall i \in [1,n]
\end{aligned}
\end{equation}
where $\boldsymbol{y}$ is the ground truth, $\boldsymbol{\hat{y}}$ is the prediction from DRS network, and $\ell_{i}$ is the $i^{th}$ candidate loss function. The continuous loss weights $\boldsymbol{\alpha} = [\alpha_1, \alpha_2, \cdots, \alpha_n]$ measure the candidates' contributions in the final loss function $\mathcal{L}$. However, this method relies on exhaustively or manually search of loss weights from a large search space, which is extremely costly. Also, this soft fusing strategy cannot completely eliminate the impact of suboptimal candidate loss functions on the final loss function $\mathcal{L}$, thus, a hard selection method is desired. However, hard selection usually leads to the training framework not end-to-end differentiable. 

Reinforcement learning (RL) is a potential solution to tackle the hard selection problem. However, since the RL is generally built upon the Markov decision process, it utilizes temporal-difference to make sequential actions. Consequently, the agent can only receive the reward until the optimal loss function is selected and the DRS is evaluated. In other words, the temporal-difference setting can suffer from delayed rewards. To address this issue, we introduce the Gumbel-softmax operation to simulate the hard selection over candidate loss functions, where the non-differentiable sampling is approximated from a categorical distribution based on a differentiable sampling from the Gumbel-softmax distribution~\cite{jang2016categorical}. 

Given the continuous loss weights $[\alpha_1, \cdots, \alpha_n]$ over candidate loss functions, we can draw a hard selection $z$ through the Gumbel-max trick~\cite{gumbel1948statistical} as: 
\begin{equation}
\begin{aligned}
z=& \;\text{one\_hot} \left(\arg \max_{i\in[1,n]}\left[\log\alpha_i+g_{i}\right]\right)\\
\end{aligned}
\end{equation}
%\end{large}
where $g_{i}=-\log \left(-\log \left(u_{i}\right)\right)$ and $u_{i} \sim Uniform (0,1)$.
The independent and identically distributed (i.i.d) \textit{gumbel noises} $\{g_{i}\}$ disturb the $\{\log\alpha_i\}$ terms. Also, they make the $\arg \max$ operation equivalent to drawing a sample from loss weights $\alpha_1, \cdots, \alpha_n$. However, because of the $\arg \max$ operation, this sampling method is non-differentiable. We tackle this problem by straight-through Gumbel-softmax~\cite{jang2016categorical}, which leverages a softmax function as a differentiable approximation to the $\arg \max$ operation: 
%\begin{large}
\begin{equation}
\label{equ:p_i}
p_i=\frac{\exp \left(\left({\log \left(\alpha_i\right)+g_{i}}\right)/{\tau}\right)}{\sum_{j=1}^{n} \exp \left(\left({\log \left(\alpha_j\right)+g_{j}}\right)/{\tau}\right)}, \quad \forall i \in [1,n] 
\end{equation}
%\end{large}
\noindent where $p_i$ is the probability of selecting the $i^{th}$ candidate loss function. The temperature parameter $\tau$ is introduced to manage the smoothness of the Gumbel-softmax operation's output. Specifically, the output approaches a one-hot vector if $\tau$ is closer to zero. Then the final loss function $\mathcal{L}$ can be reformulated as: 
\begin{equation}
\begin{aligned}
&\mathcal{L}(\boldsymbol{y}, \boldsymbol{\hat{y}}; \boldsymbol{p})=\sum_{i=1}^{n} p_{i}\cdot \ell_{i}(\boldsymbol{y}, \boldsymbol{\hat{y}})
\end{aligned}
\end{equation}

In conclusion, the loss function search process becomes end-to-end differentiable by introducing the Gumbel-softmax operation with a similar hard selection performance. Next, we will discuss how to generate data example-level loss weights $[\alpha_1, \cdots, \alpha_n]$.

%\vspace{-1mm}
\subsection{The Controller Network}
As in Eq. (\ref{eq:overall}), we suppose that $[\alpha_1, \cdots, \alpha_n]$ are the original (continuous) class probabilities over $n$ candidate loss functions before the Gumbel-softmax operation. This assumption aims to learn a set of unified and static probabilities over the candidate loss functions. However, the environment of real-world commercial recommendation platforms is always non-stationary, and different user-item interaction examples have varying convergence behaviors. This cannot be handled by unified and static probabilities, resulting in suboptimal model performance, generalizability and transferability. 

We propose a controller network to address this challenge, which learns to generate original class probabilities for each data example. Motivated by curriculum learning \cite{bengio2009curriculum,jiang2014self}, the original class probabilities should be generated according to the ground truth labels $\boldsymbol{y}$ and the output of DRS network $\boldsymbol{\hat{y}}$. Therefore, the input of the controller network is a mini-batch $\boldsymbol{(y, \hat{y})}$, followed by the MLP layer with several fully-connected layers like Eq. (\ref{eq:mlp}). Afterwards, the controller's output layer generates continuous class probabilities $[\alpha_1^b, \cdots, \alpha_n^b]\;\forall b \in [1,B]$ for each data example in the mini-batch via a standard \textit{softmax} activation, where $B$ is the size of mini-batch. In other word, each data example has individual probabilities. The controller can enhance the recommendation quality, model generalizability and transferability, which is validated by the extensive experiments. 

%\vspace{-1.2mm}
\subsection{An Optimization Method}
\label{sec:Optimization}
In above subsections, we formulate the loss function search as an architectural optimization problem and introduce the Gumbel-softmax that makes the framework end-to-end differentiable. Now, we discuss the optimization for the AutoLoss framework. 

In AutoLoss, the parameters to be optimized are from two networks. We denote the main DRS network's parameters as $\boldsymbol{W}$, and the controller network's parameters as $\boldsymbol{V}$. Note that $\boldsymbol{p}$ are directly generated by the Gumbel-softmax operation based on the controller's output $\boldsymbol{\alpha}$ as in Eq. (\ref{equ:p_i}). Inspired by automated machine learning techniques~\cite{pham2018efficient}, $\boldsymbol{W}$ and $\boldsymbol{V}$ should not be updated on the same training data batch like traditional supervised learning methods. This is because the optimization of them is highly dependent on each other. As a result, updating $\boldsymbol{W}$ and $\boldsymbol{V}$ on the same training batch can lead to the model over-fitting on the training examples.

\begin{algorithm}[t]
	\caption{\label{alg:DARTS} An Optimization Algorithm for AutoLoss via DARTS.}
	\raggedright
	{\bf Input}: features $\boldsymbol{x}$ and ground-truth labels $\boldsymbol{y}$\\
	{\bf Output}: well-learned parameters $\boldsymbol{W}^*$ and  $\boldsymbol{V}^*$\\
	\begin{algorithmic} [1]
		\WHILE{not converged}
		\STATE Sample a mini-batch of validation data examples 
		\STATE Estimate the approximation of $\boldsymbol{W}^*(\boldsymbol{V})$ via Eq.(\ref{equ:approximation})
		\STATE Update $\boldsymbol{V}$ by descending $\nabla_{\boldsymbol{V}} \;\mathcal{L}_{val} \big(\boldsymbol{W}^*(\boldsymbol{V}),\boldsymbol{V}\big)$
		\STATE Sample a mini-batch of training data examples
		\STATE Update $\boldsymbol{W}$ by descending $\nabla_{\boldsymbol{W}}\mathcal{L}_{train} (\boldsymbol{W}, \boldsymbol{V})$
		\ENDWHILE
	\end{algorithmic}
\end{algorithm}
%\end{large} 

According to the end-to-end differentiable property of AutoLoss, we update $\boldsymbol{W}$ and $\boldsymbol{V}$ through gradient descent utilizing the differentiable architecture search (DARTS) techniques~\cite{liu2018darts}. To be specific, $\boldsymbol{W}$ and $\boldsymbol{V}$ are alternately updated on training and validation batches by minimizing the training loss $\mathcal{L}_{train}$ and validation loss $\mathcal{L}_{val}$, respectively. This forms a bi-level optimization problem~\cite{pham2018efficient}, 
where controller parameters $\boldsymbol{V}$ and DRS parameters $\boldsymbol{W}$ are considered as the upper- and lower-level variables: 
%\begin{large}
\begin{equation}
\begin{aligned}
\label{equ:bilevel}
\min_{\boldsymbol{V}} \; &\mathcal{L}_{val} \big(\boldsymbol{W}^*(\boldsymbol{V}),\boldsymbol{V}\big)\\
s.t. \; & \boldsymbol{W}^*(\boldsymbol{V}) = \arg\min_{\boldsymbol{W}} \mathcal{L}_{train} (\boldsymbol{W}, \boldsymbol{V}^*)
\end{aligned}
\end{equation}
%\end{large}
where directly optimizing $\boldsymbol{V}$ thoroughly via Eq.(\ref{equ:bilevel}) is intractable since the inner optimization of $\boldsymbol{W}$ is extremely costly. To tackle this issue, we use an approximation scheme for the inner optimization:
%\begin{large}
\begin{equation}
\begin{aligned}
\label{equ:approximation}
& \boldsymbol{W}^*(\boldsymbol{V})\approx \boldsymbol{W} - \xi \nabla_{\boldsymbol{W}}\mathcal{L}_{train} (\boldsymbol{W}, \boldsymbol{V})
\end{aligned}
\end{equation}
%\end{large}
\noindent where $\xi$ is the predefined learning rate. This approximation scheme estimates $\boldsymbol{W}^*(\boldsymbol{V})$ by descending only one step toward the gradient $\nabla_{\boldsymbol{W}}\mathcal{L}_{train} (\boldsymbol{W}, \boldsymbol{V})$, rather than optimizing $\boldsymbol{W}(\boldsymbol{V})$ thoroughly. To further enhance the computation efficiency, we can set $\xi=0$, i.e., the first-order approximation.

We detail the AutoLoss optimization via DARTS in Algorithm \ref{alg:DARTS}. More specifically, in each iteration, we first sample a mini-batch validation data examples of user-item interactions (line 2); next, we estimate (but do not update) $\boldsymbol{W}^*(\boldsymbol{V})$ via the approximation scheme in Eq.(\ref{equ:approximation}) (line 3); then, we update the controller parameters  $\boldsymbol{V}$ by one step based on the estimation (line 4); 
afterward, we sample a mini-batch training data examples (line 5); and finally, we update the $\boldsymbol{W}$ via descending $\nabla_{\boldsymbol{W}}\mathcal{L}_{train} (\boldsymbol{W}, \boldsymbol{V})$ by one step (line 6).

%% file: 3Experiments.tex
\section{Experiment}
\label{sec:experiment}
This section will conduct extensive experiments using various datasets to evaluate the effectiveness of AutoLoss. We first introduce the experimental settings, then compare AutoLoss with representative baselines, and finally conduct model component and transferability analysis.

\subsection{Datasets}
\label{sec:Datasets}
We evaluate our model on two datasets, including Criteo and ML-20m. Below we introduce these datasets and more statistics about them can be found in Table~\ref{table:statistics}. 

\textbf{Criteo}\footnote{https://www.kaggle.com/c/criteo-display-ad-challenge/}: It is a real-world commercial dataset to assess click-through rate prediction models for online ads. It consists of 45 million data examples, i.e., users' click records on displayed ads. Each example contains $m=39$ anonymous feature fields, where 13 fields are numerical and 26 fields are categorical. 13 numerical fields are converted into categorical features through bucketing. %The features in a certain field appearing less than 20 times are set as a dummy feature "other" for dimensionality reduction.

\textbf{ML-20m}\footnote{https://grouplens.org/datasets/movielens/20m/}: This is a benchmark dataset to evaluate recommendation algorithms, which contains 20 million users' 5-star ratings on movies. The dataset includes 27,278 movies and 138,493 users, i.e., $m=2$ feature fields, where each user has at least 20 ratings.

%\vspace{-2mm}
\subsection{Evaluation Metrics}
\label{sec:metrics}
AutoLoss is general for many recommendation tasks. To evaluate its effectiveness, we conduct \textit{binary classification} (i.e., click-through rate prediction) on Criteo, and \textit{multi-class classification} (i.e., 5-star ratings) on ML-20m.
%,  \textit{regression} (i.e., dwell time prediction) on Douyin. 
The two classification experiments are evaluated by AUC\footnote{We evaluate the AUC for multiclass classification in a one-vs-rest manner.} and Logloss, where higher AUC or lower Logloss mean better performance. It is worth noting that slightly higher AUC and lower Logloss at 0.001-level are considered significant in recommendations~\cite{guo2017deepfm}.

\begin{table}[]
	\caption{Statistics of the datasets.}
	\label{table:statistics}
	\begin{tabular}{@{}|c|c|c|@{}}
		\toprule[1pt]
		Data & Criteo & ML-20m  \\ \midrule
		\# Interactions  & 45,840,617 & 20,000,263  \\
		\# Feature Fields  & 39 & 2  \\
		\# Feature Values  & 1,086,810 & 165,771  \\
		\# Behavior & click or not & rating 1$\sim$5 \\ \bottomrule[1pt]
	\end{tabular}
	\vspace{-2.1mm}
\end{table}

%\vspace{-2.1mm}
\subsection{Implementation}
\label{sec:architecture}
We implement AutoLoss based on a public library\footnote{https://github.com/rixwew/pytorch-fm}, which contains 16 representative recommendation models. We develop AutoLoss as an independent class, so we can easily apply our framework for all these models. In this paper, we only show the results on DeepFM~\cite{guo2017deepfm} and IPNN~\cite{qu2016product} due to the page limitation. To be specific, AutoLoss framework mainly contains two networks, i.e., the DRS network and the controller network. 

\begin{table*}[]
	\caption{Performance comparison of different loss function search methods.}
	\label{table:result1}
	\begin{tabular}{@{}|c|c|c|ccccccccc|@{}}
		\toprule[1pt]
		\multirow{2}{*}{Dataset} & \multirow{2}{*}{Model} & \multirow{2}{*}{Metric} & \multicolumn{9}{c|}{Methods} \\ \cmidrule(l){4-12} 
		&  &  & Focal & KL & Hinge & CE & MeLU & BOHB & DARTS & SLF & AutoLoss \\ \midrule
		\multirow{2}{*}{Criteo} & \multirow{2}{*}{DeepFM} 
		& AUC $\uparrow$ 				& 0.8046 & 0.8042 & 0.8049 & 0.8056 & 0.8063 & 0.8065 & 0.8067 & 0.8081 & \textbf{0.8092*} \\
		&  & Logloss $\downarrow$& 0.4466 & 0.4469 & 0.4463 & 0.4457 & 0.4436 & 0.4435 & 0.4433 & 0.4426 & \textbf{0.4416*} \\ \midrule
		
		\multirow{2}{*}{Criteo} & \multirow{2}{*}{IPNN} 
		& AUC $\uparrow$				& 0.8077 & 0.8072 & 0.8079 & 0.8085 & 0.8090 & 0.8092 & 0.8093 & 0.8098 & \textbf{0.8108*} \\
		&  & Logloss $\downarrow$& 0.4435 & 0.4437 & 0.4432 & 0.4428 & 0.4423 & 0.4422 & 0.4423 & 0.4418 & \textbf{0.4409*} \\ \midrule
		
		\multirow{2}{*}{ML-20m} & \multirow{2}{*}{DeepFM} 
		& AUC $\uparrow$ 				& 0.7681 & 0.7682 & 0.7685 & 0.7692 & 0.7695 & 0.7695 & 0.7696 & 0.7705 & \textbf{0.7717*} \\
		&  & Logloss $\downarrow$& 1.2320 & 1.2317 & 1.2316 & 1.2310 & 1.2307 & 1.2305 & 1.2305 & 1.2299 & \textbf{1.2288*} \\ \midrule
		
		\multirow{2}{*}{ML-20m} & \multirow{2}{*}{IPNN} 
		& AUC $\uparrow$				& 0.7721 & 0.7722 & 0.7725 & 0.7733 & 0.7735 & 0.7734 & 0.7736 & 0.7745 & \textbf{0.7756*} \\
		&  & Logloss $\downarrow$& 1.2270 & 1.2269 & 1.2266 & 1.2260 & 1.2256 & 1.2257 & 1.2255 & 1.2249 & \textbf{1.2236*} \\ \bottomrule[1pt]
	\end{tabular}
	\\``\textbf{{\Large *}}'' indicates the statistically significant improvements (i.e., two-sided t-test with $p<0.05$) over the best baseline.
	\\ $\uparrow$: the higher the better; $\downarrow$: the lower the better.
	%	\vspace{-3mm}
\end{table*}

For the DRS network, 
(a) \textit{Embedding layer}: we set the embedding size as 16 following the existing works~\cite{zhu2020fuxictr}.
(b) \textit{Interaction layer}: we leverage factorization machine and inner product network to capture the interactions among feature fields for DeepFM and IPNN, respectively.
(c) \textit{MLP layer}: we have two fully-connected layers, and the layer size is 128. We also employ batch normalization, dropout ($rate=0.2$) and ReLU activation for both layers. 
(d) \textit{Output layer}: original DeepFM and IPNN are designed for click-through rate prediction, which use \textit{sigmoid} activation for negative log-likelihood function. To fit the 5-class classification task on ML-20m, we modify the output layer correspondingly. i.e., the output layer is 5-dimensional with \textit{softmax} activation.

For the controller network, (a) \textit{Input layer}: the inputs are the ground truth labels $\boldsymbol{y}$ and the predictions $\boldsymbol{\hat{y}}$ from DRS network. 
(b) \textit{MLP layer}: we also use two fully-connected layers with the layer size 128, batch normalization, dropout ($rate=0.2$) and ReLU activation. (3) \textit{Output layer}: the controller network will output continuous loss probabilities $\boldsymbol{\alpha}$ with \textit{softmax} activation, whose dimension equals to the number of candidate loss functions.

For other hyper-parameters, (a) \textit{Gumbel-softmax}: we use an annealing scheme for temperature $\tau=\max (0.01, 1 - 0.00005 \cdot t)$, where $t$ is the training step. (b) \textit{Optimization}: we set the learning rate as $0.001$ for updating both DRS network and controller network with Adam optimizer and batch-size 2000. 
(c) We select the hyper-parameters of the AutoLoss framework via cross-validation, and we also do parameter-tuning for baselines correspondingly for a fair comparison.

\subsection{Overall Performance Comparison}
\label{sec:RQ1}
AutoLoss is compared with the following loss function design and search methods:
\begin{itemize}[leftmargin=*]
	\item Fixed loss function: the first group of baselines leverages a predefined and fixed loss function. We utilize Focal loss, KL divergence, Hinge loss and cross-entropy (CE) loss for both classification tasks.
	\item Fixed weights over loss functions: this group of baselines aims to learn fixed weights over the loss functions in the first group, without considering the difference among data examples. In this group, we use the meta-learning method MeLU~\cite{lee2019melu}, as well as automated machine learning methods BOHB~\cite{falkner2018bohb} and DARTS~\cite{liu2018darts}.
	\item Data example-wise loss weights: this group learns to assign different loss weights for different data examples according to their convergence behaviors. One existing work, stochastic loss function (SLF)~\cite{liu2020stochastic}, belongs to this group.
\end{itemize}

The overall performance is shown in Table~\ref{table:result1}. It can be observed that: 
%\begin{itemize}[leftmargin=*]
(i) The first group of baselines achieves the worst recommendation performance in both recommendation tasks. Their optimizations are based on predefined and fixed loss functions during the training stage. This result demonstrates that leveraging a predefined and fixed loss function throughout can downgrade the recommendation quality.
(ii) The methods in the second group outperform those in the first group. These methods try to learn weights over candidate loss functions according to their contributions to the optimization, and then combine them in a weighted sum manner. This validates that incorporating multiple loss functions in optimization can enhance the performance of deep recommender systems.  
(iii) The second group performs worse than the SLF, since the weights they learned are unified and static, which completely overlooks the various behaviors among different data examples. Therefore, SLF performs better by taking this factor into account.
(iv) The decision network of SLF is optimized on the same training batch with the main DRS network via  back-propagation, which can lead to over-fitting on the training batch. AutoLoss updates the DRS network on the training batch while updating the controller on the validation batch, which improves the model generalizability and results in better recommendation performance.
%\end{itemize}

To summarize, AutoLoss achieves significantly better performance than state-of-the-art baselines on both datasets and tasks, which demonstrates its effectiveness.

\subsection{Transferability Study}
\label{sec:RQ2}
In this subsection, we study the transferability of the controller. Specifically, we want to investigate (i) whether the controller trained with one DRS model can be applied to other DRS models; and (ii) whether the controller learned on one dataset can be directly used on other datasets.

\begin{figure}[t]
	\centering
	%	\hspace*{-0.6cm}
	{\subfigure{\includegraphics[width=0.327\linewidth]{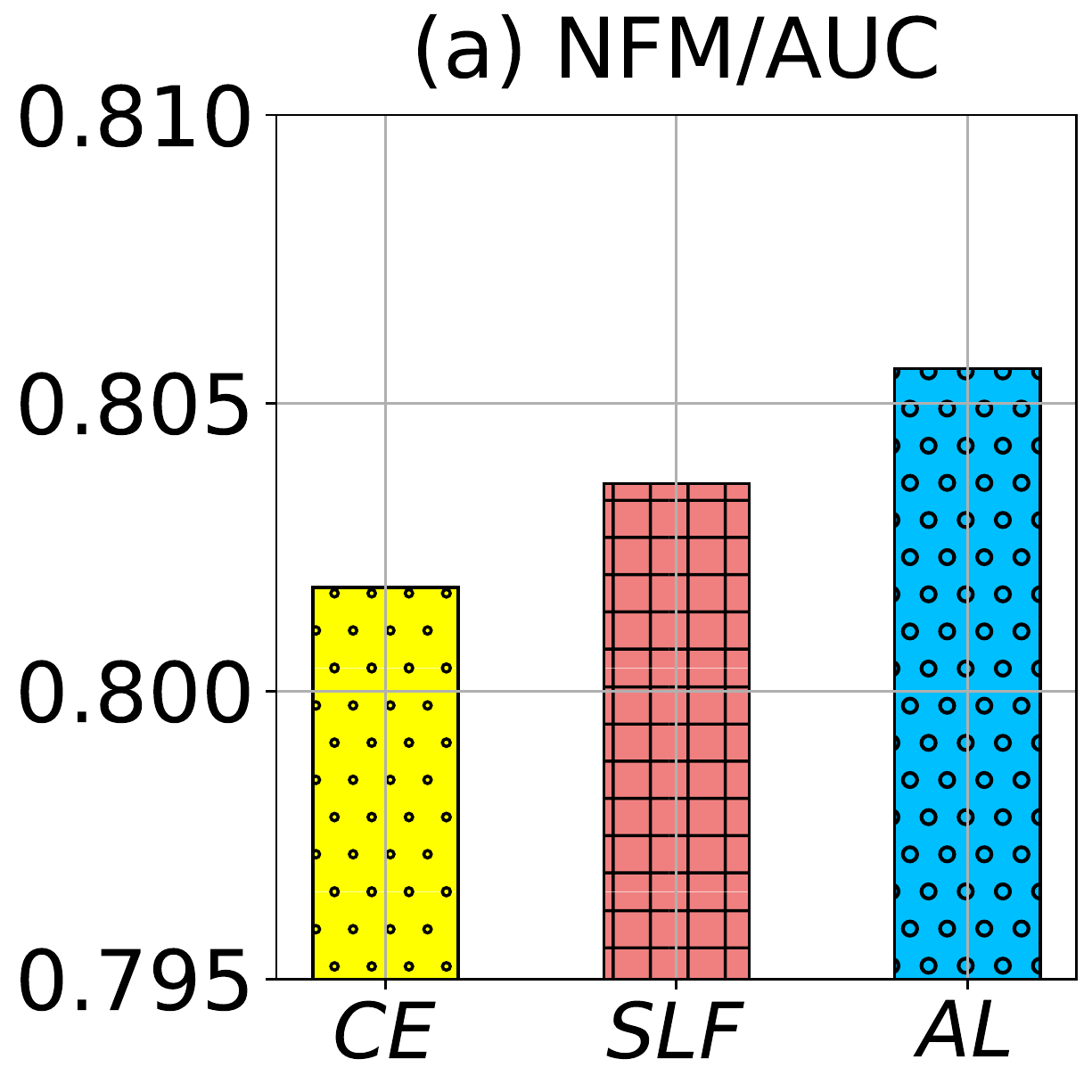}}}
	{\subfigure{\includegraphics[width=0.327\linewidth]{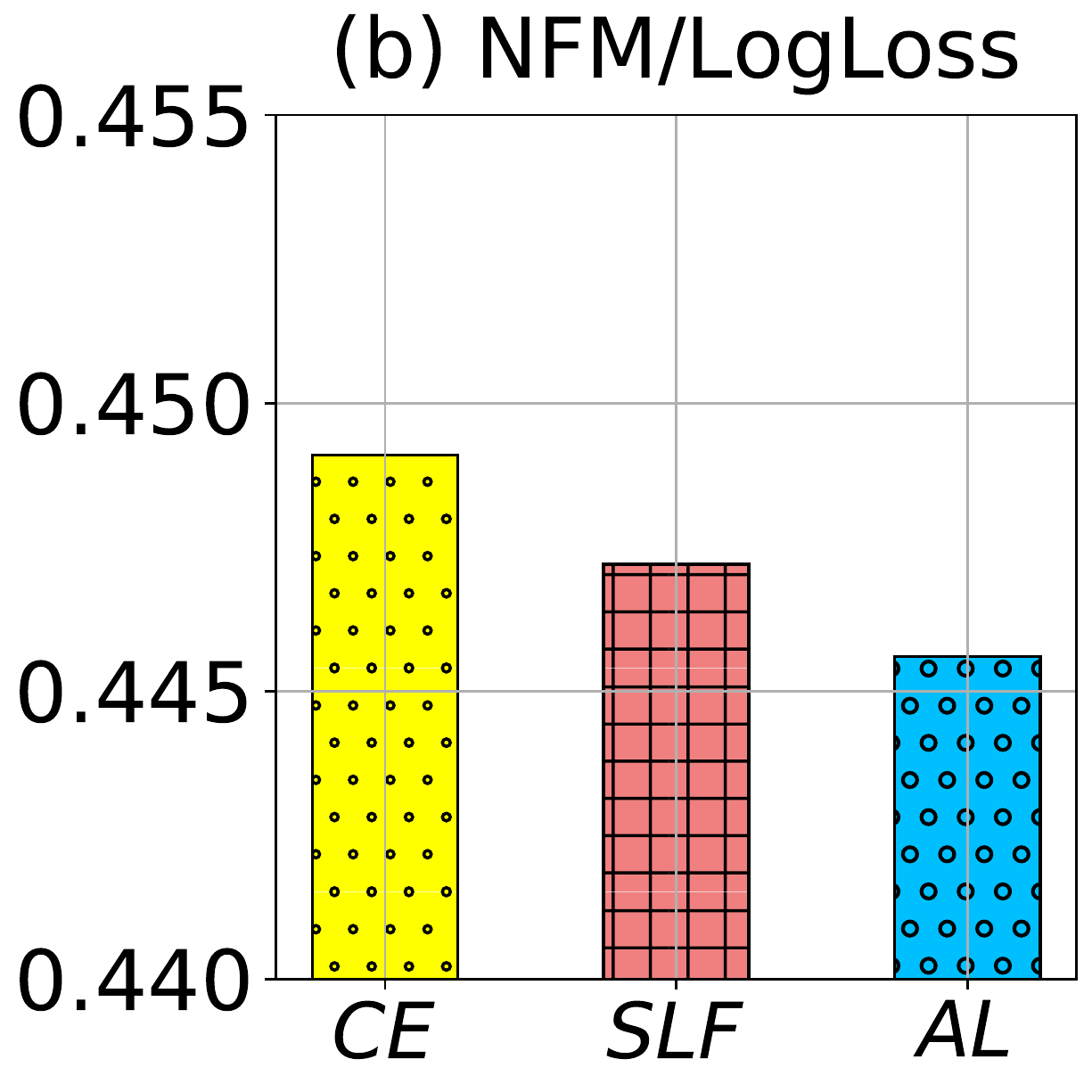}}}
	{\subfigure{\includegraphics[width=0.327\linewidth]{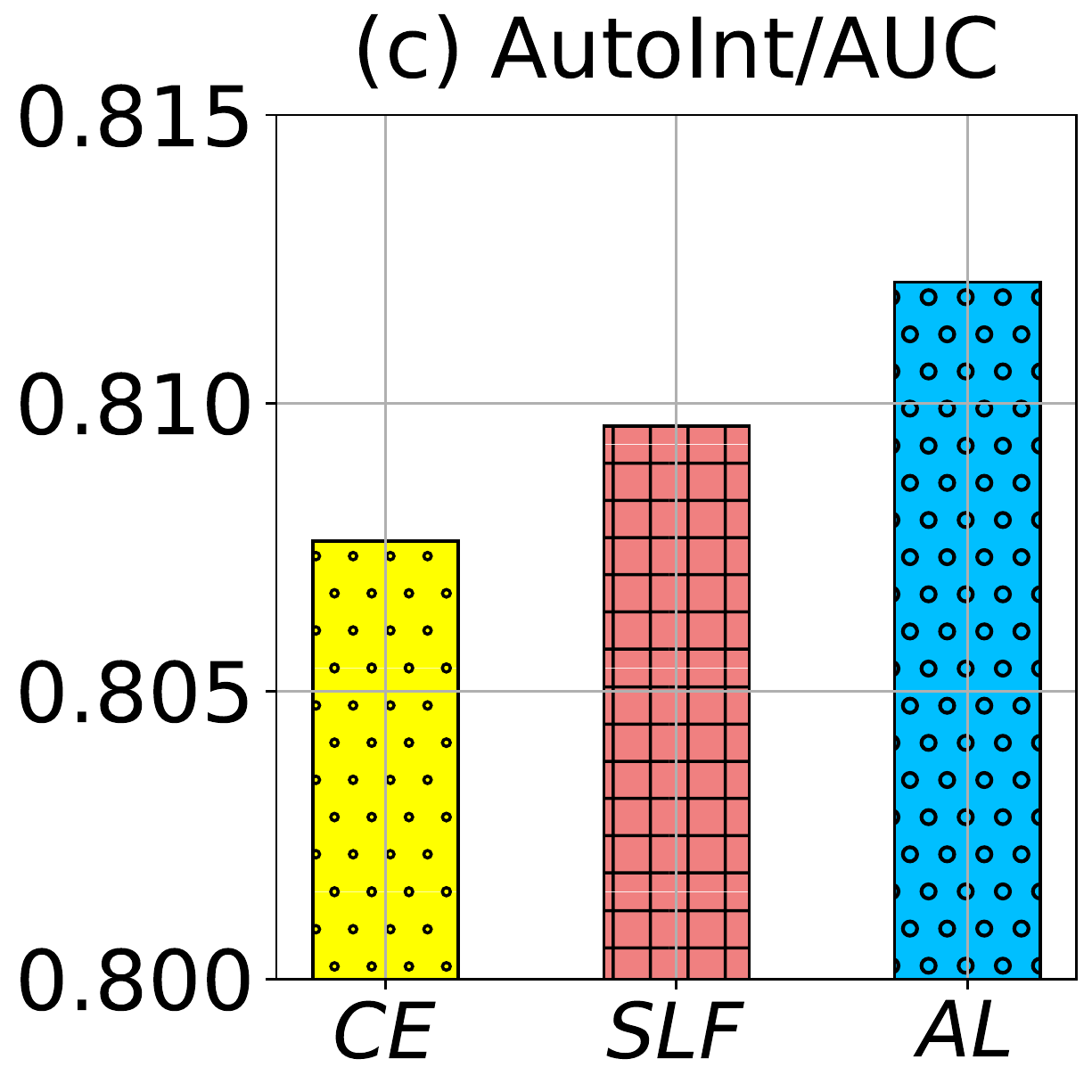}}}
	%	\hspace*{-0.6cm}
	{\subfigure{\includegraphics[width=0.327\linewidth]{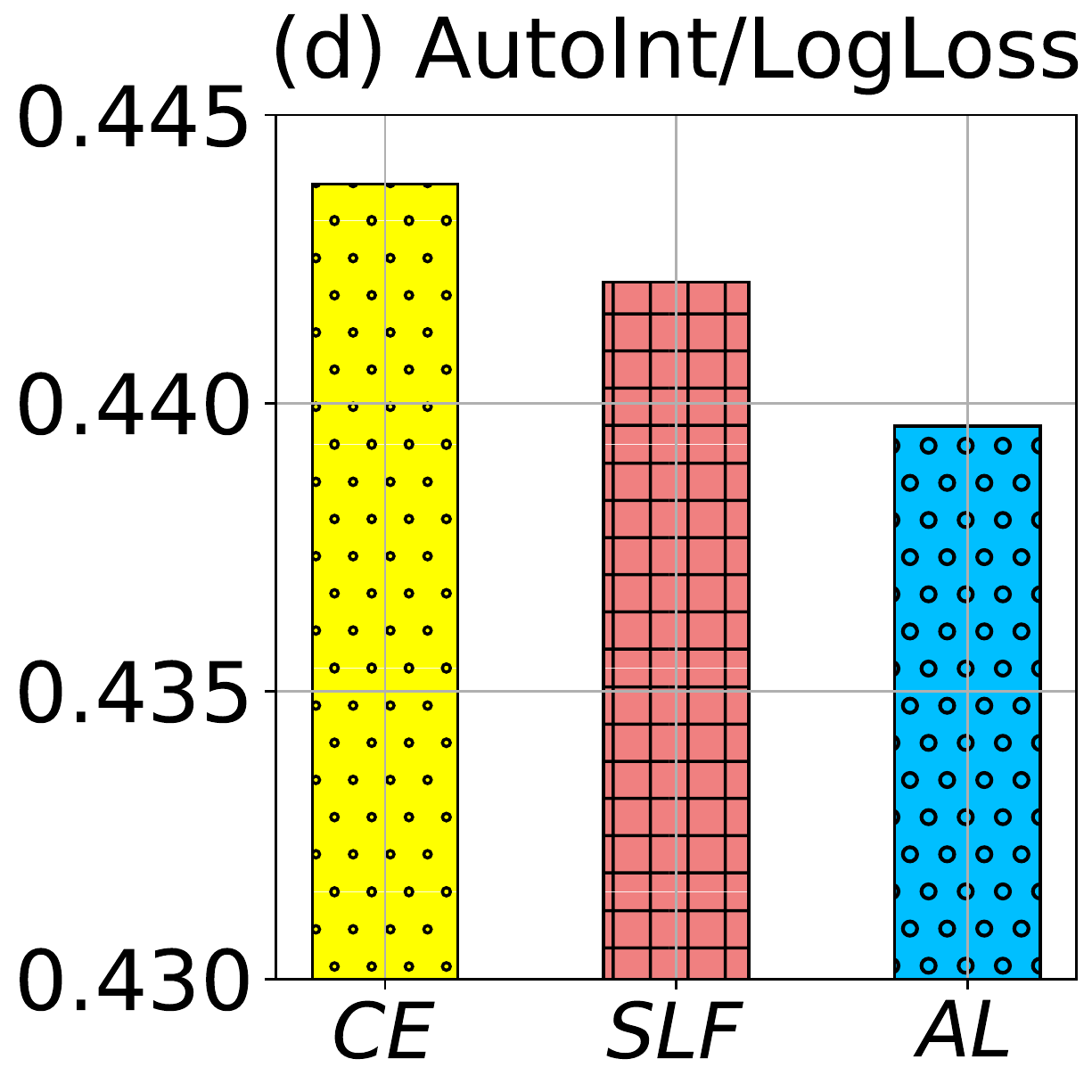}}}
	{\subfigure{\includegraphics[width=0.327\linewidth]{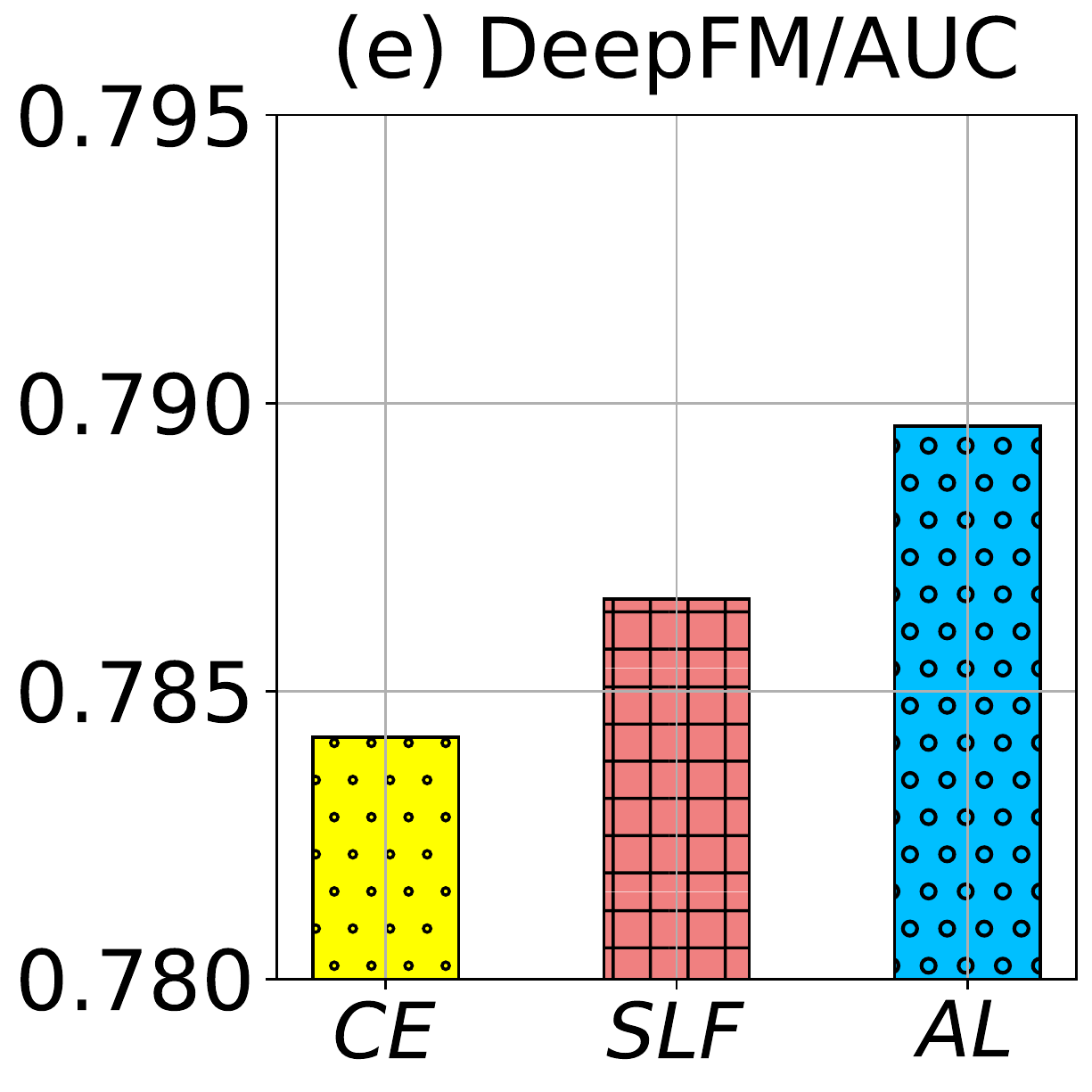}}}
	{\subfigure{\includegraphics[width=0.327\linewidth]{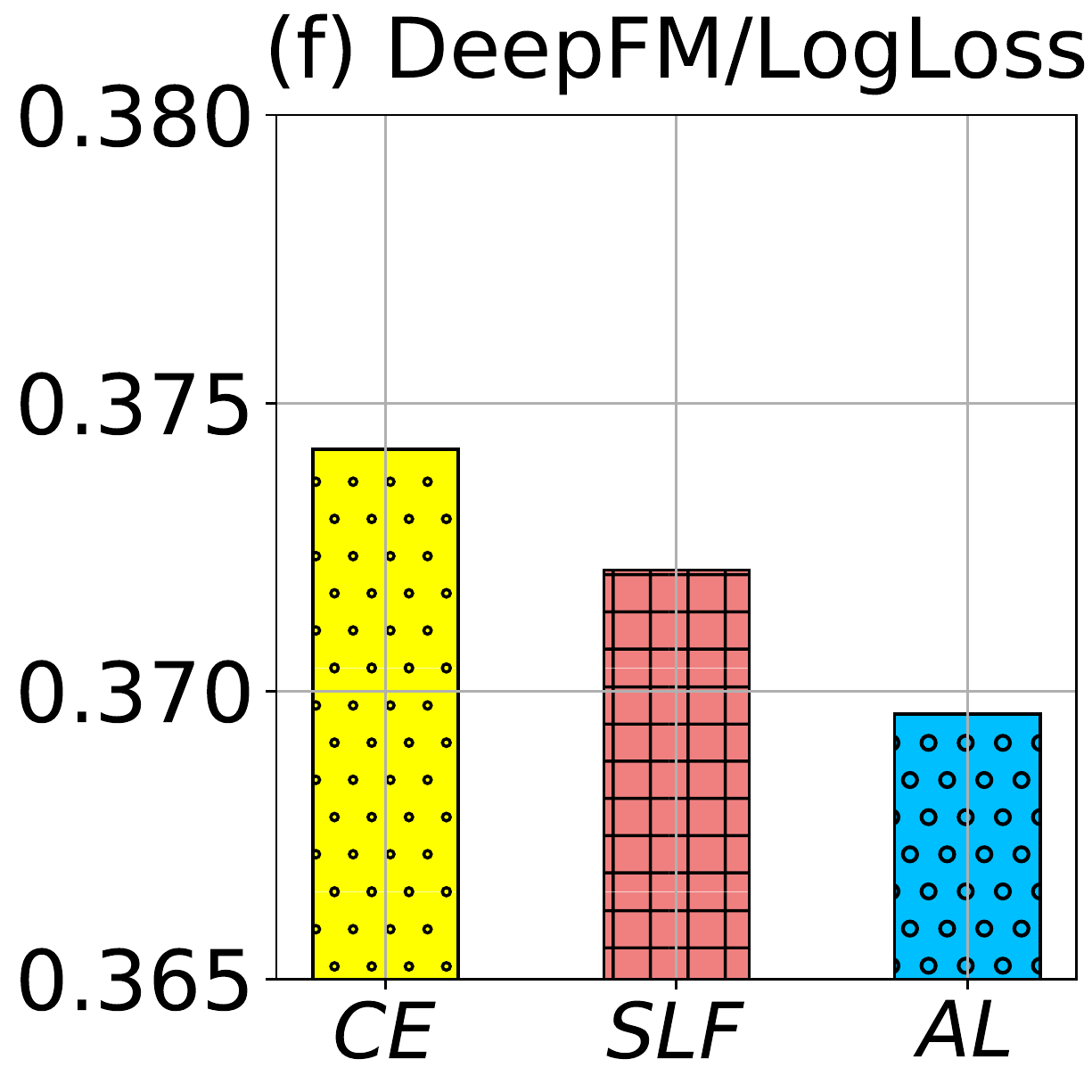}}}
	\caption{Transferability study results.}\label{fig:Fig3}
	\vspace{-4.9mm}
\end{figure}

To study the transferability across different DRS models, we leverage the controller trained via DeepFM and AutoLoss on Criteo, fix its parameters and apply it to train NFM~\cite{he2017neural} and AutoInt~\cite{song2019autoint} on Criteo. The results are demonstrated in Figure~\ref{fig:Fig3} (a)-(d), where (i) ``${CE}$'' means that we directly train the new DRS model via  minimizing the cross-entropy (CE) loss, which is the best single and fixed loss function in Table~\ref{table:result1}; (ii) ``${SLF}$'' is that we use the controller upon DeepFM and SLF, which is the best baseline in Table~\ref{table:result1}; and (iii) ``${AL}$'' denotes that we use the controller based on DeepFM and AutoLoss. From the figures, we can observe that ${SLF}$ performs superior to ${CE}$, which indicates that a pre-trained controller can improve other DRS models' training performance. More importantly, ${AL}$ outperforms ${SLF}$, which validates AutoLoss's better transferability across different DRS models.

To study the transferability between different datasets, we train a controller upon Criteo dataset with DeepFM and AutoLoss, fix its parameters and apply it to train a new DeepFM on the Avazu dataset\footnote{Avazu is another benchmark dataset for CTR prediction, which contains 40 million user clicking behaviors in 11 days with $M=22$ categorical feature fields. https://www.kaggle.com/c/avazu-ctr-prediction/}, i.e., ``${AL}$''. Also, we denote that (i) ``${CE}$'': DeepFM is directly optimized by minimizing cross-entropy (CE) loss on Avazu dataset; and (ii) ``${SLF}$'': DeepFM is optimized on the new dataset with the assistance of a controller pre-trained with DeepFM+SLF on Criteo. In Figure~\ref{fig:Fig3} (e)-(f), ${AL}$ shows superior performance over ${CE}$ and ${SLF}$, which proves its better transferability between different datasets.

In summary, AutoLoss has better transferability across different DRS models and different recommendation datasets, which improves its usability in real-world recommender systems.

\subsection{Impact of Model Components}
\label{sec:RQ3}

In this subsection, in order to understand the contributions of important model components of AutoLoss, we systematically eliminate each component and define the following variants: 
\begin{itemize}[leftmargin=*]
	\item \textbf{AL-1}: This variant aims to assess the contribution of the controller. Thus, we assign equivalent weights on four candidate loss functions, i.e., $[0.25,0.25,0.25,0.25]$.
	\item \textbf{AL-2}: In this variant, we eliminate the Gumbel-softmax operation, and directly use the controller's output, i.e., the continuous loss probabilities $\boldsymbol{\alpha}$ from standard \textit{softmax} activation, which aims to evaluate the impact of Gumbel-softmax.
\end{itemize}

\begin{table}[t]
	\caption{ Impact of model components.}
	\label{table:result2}
	\begin{tabular}{@{}|c|c|c|ccc|@{}}
		\toprule[1pt]
		\multirow{2}{*}{Dataset} & \multirow{2}{*}{Model} & \multirow{2}{*}{Metric} & \multicolumn{3}{c|}{Methods} \\ \cmidrule(l){4-6} 
		&  &  & AL-1 & AL-2 & AutoLoss \\ \midrule
		\multirow{2}{*}{Criteo} & \multirow{2}{*}{DeepFM} 
		& AUC $\uparrow$& 0.8052 & 0.8083 & \textbf{0.8092*} \\
		&  & Logloss $\downarrow$& 0.4460 & 0.4422 & \textbf{0.4416*} \\ \midrule
		
		\multirow{2}{*}{Criteo} & \multirow{2}{*}{IPNN} 
		& AUC $\uparrow$& 0.8081 & 0.8102 & \textbf{0.8108*} \\
		&  & Logloss $\downarrow$& 0.4431 & 0.4416 & \textbf{0.4409*} \\ \bottomrule[1pt]
	\end{tabular}
	\\``\textbf{{\Large *}}'' indicates the statistically significant improvements (i.e., two-sided t-test with $p<0.05$) over the best baseline.
	\\ $\uparrow$: the higher the better; $\downarrow$: the lower the better.
		\vspace{-3mm}
\end{table}

The results on the Criteo dataset are shown in Table \ref{table:result2}. First, AL-1 has worse performance than AutoLoss, which validates the necessity to introduce the controller network. It is noteworthy that, AL-1 performs worse than all loss function search methods, and even the fixed cross-entropy (CE) loss in Table~\ref{table:result1}, which indicates that equally incorporating all candidate loss functions cannot guarantee better performance. Second, AutoLoss outperforms AL-2. The main reason is that AL-2 always generates gradients based on all the loss functions, which introduces some noisy gradients from the suboptimal candidate loss functions. In contrast, AutoLoss can obtain appropriate gradients by filtering out suboptimal loss functions via Gumbel-softmax, which enhances the model robustness.

\begin{figure}[t]
	\centering
	%	\hspace*{-1.25mm}
	{\subfigure{\includegraphics[width=0.327\linewidth]{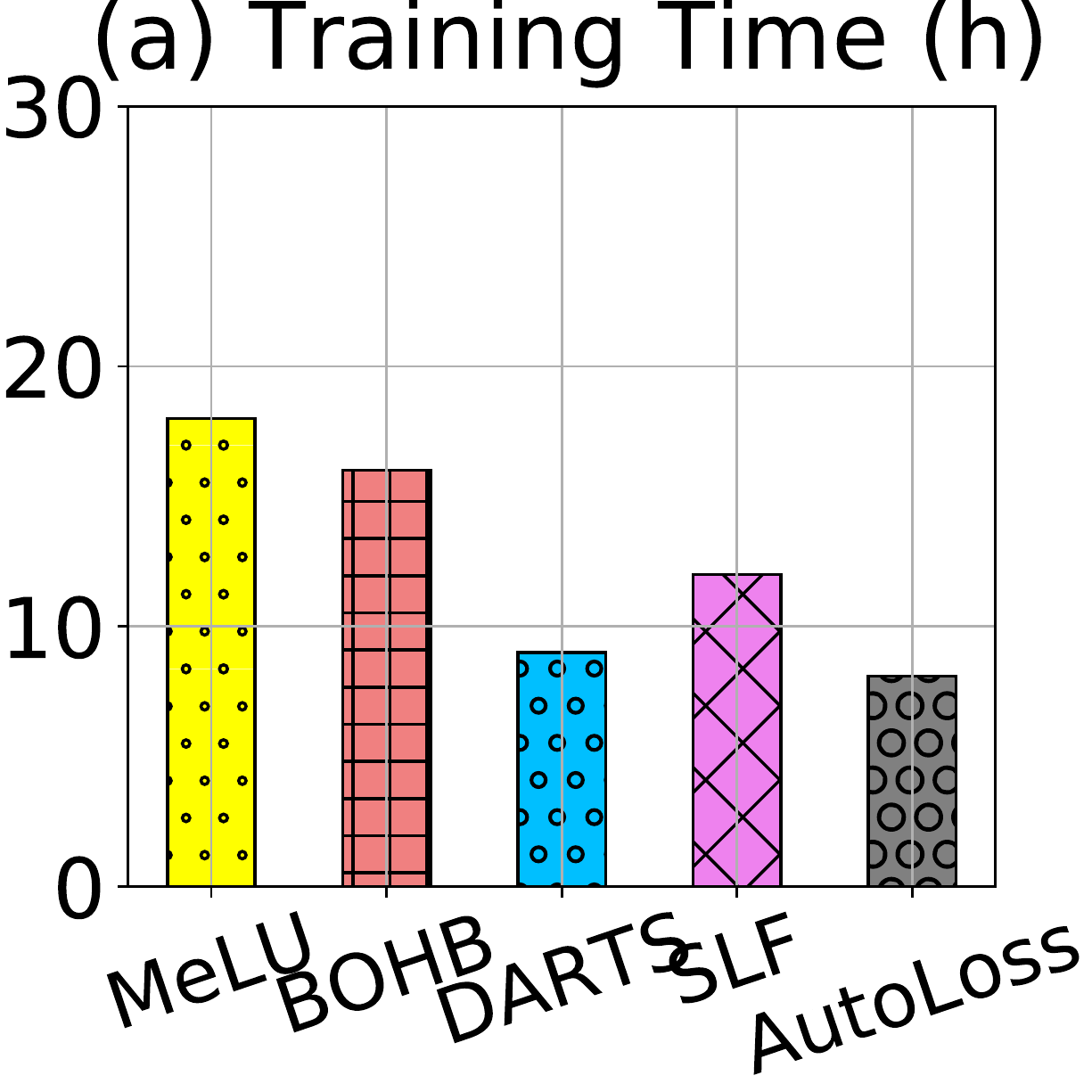}}}
	{\subfigure{\includegraphics[width=0.327\linewidth]{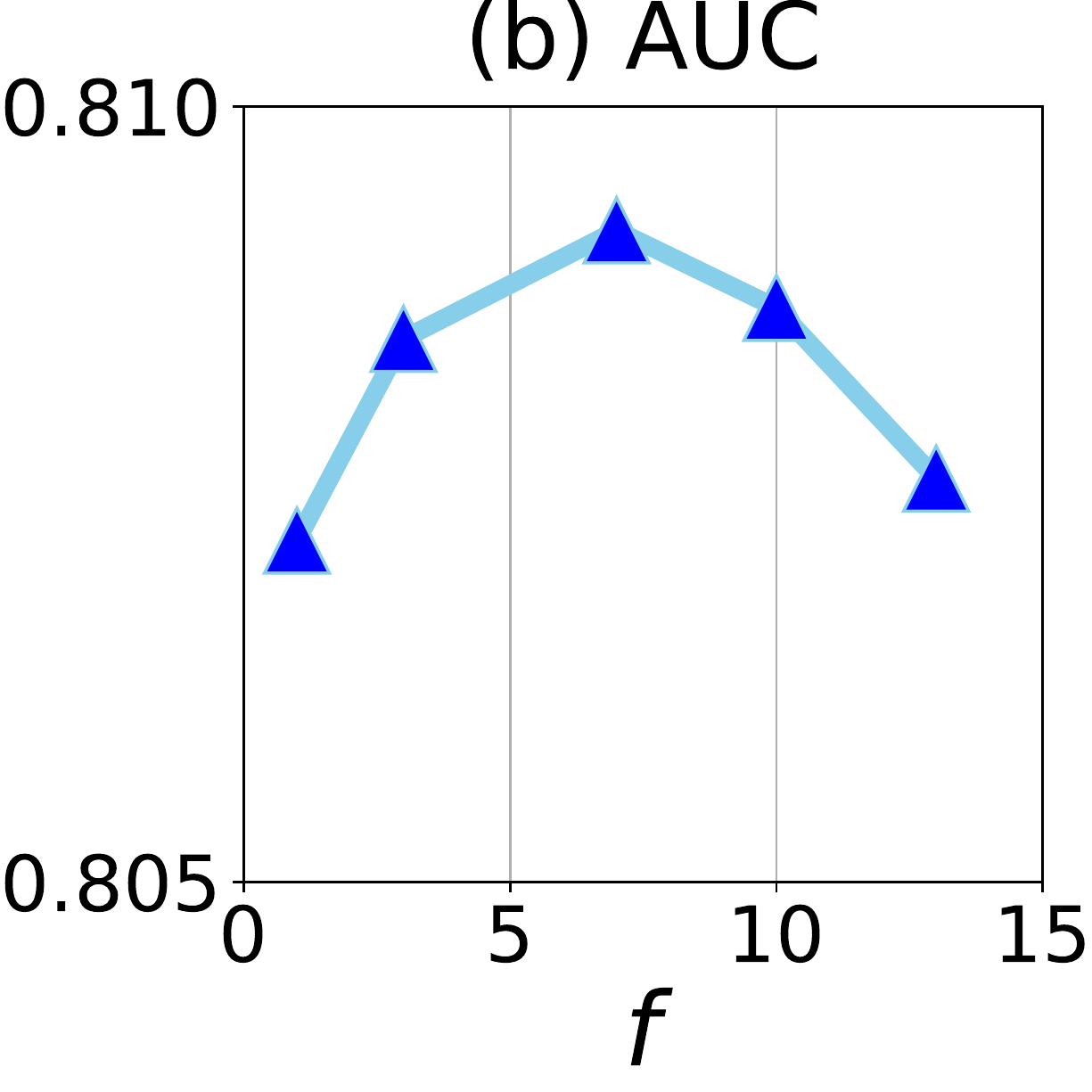}}}
	{\subfigure{\includegraphics[width=0.327\linewidth]{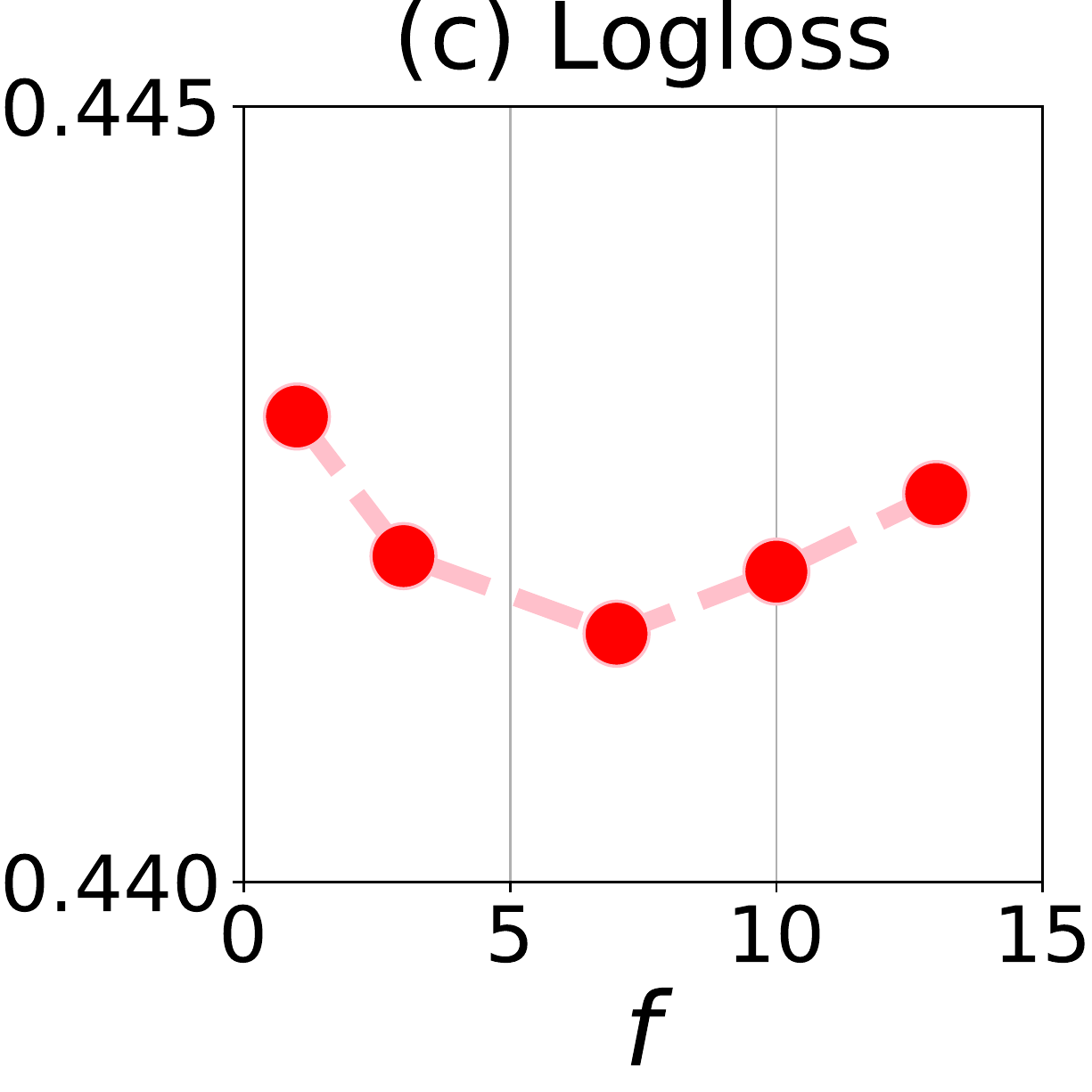}}}%\hspace*{7mm}
	\caption{Efficiency study results.}
	\label{fig:Fig5}
		\vspace{-3.3mm}
\end{figure}
\subsection{Efficiency Study}
\label{sec:RQ5}
This section compares AutoLoss's training efficiency with other loss function searching methods, which is an important metric to deploy a DRS model in real-world applications. Our experiments are based on one GeForce GTX 1060 GPU. 

The results of DeepFM on Criteo dataset are illustrated in Figure~\ref{fig:Fig5} (a). We can observe that AutoLoss achieves the fastest training speed. The reasons are two-fold. First, AutoLoss can generate the most appropriate gradients to update DRS, which increases the optimization efficiency. Second, we update the controller once after every 7 times DRS is updated, i.e., the controller updating frequency $f=7$. This trick not only reduces the training time ($\sim60\%$ in this case) with fewer computations, but also enhances the performance. In Figure~\ref{fig:Fig5} (b)-(c) where $x$-axis is $f$, we find that DeepFM performs the best when $f=7$, while updating too frequently/infrequently can lead to suboptimal AUC/Logloss. 

To summarize, AutoLoss can efficiently achieve better performance, making it easier to be launched in real-world recommender systems.

%% file: 5RelatedWork.tex
\section{Related Work}
\label{sec:related_work}
In this section, we briefly introduce the works related to our study. We first go over the latest studies in loss function search and then review works about AutoML for recommendations.

\vspace{-2.1mm}
\subsection{Loss Function Search}
The loss function plays an essential part in a deep learning framework. The choice of the loss function significantly affects the performance of the learned model. A lot of efforts have been made to design desirable loss functions for specific tasks. 
For example, in the field of image processing, \citet{rahman2016optimizing} argued that the typical cross-entropy loss for semantic segmentation shows great limitations in aligning with evaluation metrics other than global accuracy. 
\citet{ronneberger2015u,wu2016bridging} designed loss functions by taking class frequency into consideration to cater to the mIoU metric. 
\citet{caliva2019distance,qin2019basnet} designed losses with larger weights at boundary regions to improve the boundary F1 score. 
\citet{liu2016large} proposed to replace the traditional Softmax loss with large margin Softmax (L-Softmax) loss to improve feature discrimination in classification tasks. 
\citet{fan2019spherereid} used sphere Softmax loss for the person re-identification task and obtained state-of-the-art results. 
The loss functions mentioned above are all designed manually, requiring ample expert knowledge, non-trivial time, and many human efforts. 

Recently, automated loss function search draws increasing interests of researchers from various machine learning (ML) fields. 
\citet{xu2018autoloss} investigated how to automatically schedule iterative and alternate optimization processes for ML models. 
%A meta-learning framework was proposed to adaptively determine which loss function to use and which parameters to update at each optimization step. 
%The proposed framework shows its generalization ability on various typical ML tasks, such as $d$-ary quadratic regression, classification, image generation and machine translation. 
\citet{liu2020stochastic} proposed to optimize the stochastic loss function (SLF), where the loss function of an ML model was dynamically selected. 
%The loss function selection is determined by loss parameters, including a selective binary code and a weighting coefficient. 
During training, model parameters and the loss parameters are learned jointly. 
\citet{li2020auto} proposed automatically searching specific surrogate losses to improve different evaluation metrics in the image semantic segmentation task. 
\citet{jin2021automated} composed multiple self-supervised learning tasks to jointly encode multiple sources of information and produce more generalizable representations, and developed two automated frameworks to search the task weights.
Besides, \citet{li2019lfs,wang2020loss} designed search spaces for a series of existing loss functions and developed algorithms to search for the best parameters of the probability distribution for sampling loss functions. 
%However, their methods are designed exclusively for cross-entropy loss and its variants, making their methods not applicable in our tasks. 

%\vspace{-3mm}
\subsection{AutoML for Recommendation}
AutoML techniques are now widely used to automatically design deep recommendation systems. Previous works mainly focused on the design of the embedding layer and the selection of feature interaction patterns.

In terms of the embedding layer, \citet{joglekar2020neural,zhao2020memory,liu2021learnable} proposed novel methods to automatically select the best embedding size for different feature fields in a recommendation system. \citet{zhao2020autoemb,liu2020automated} proposed to dynamically search embedding sizes for users and items based on their popularity in the streaming setting. 
Similarly, \citet{ginart2019mixed} proposed to use mixed dimension embeddings for users and items based on their query frequency. 
\citet{kang2020learning} proposed a multi-granular quantized embeddings (MGQE) technique to learn impact embeddings for infrequent items. 
\citet{cheng2020differentiable} proposed to perform embedding dimension selection with a soft selection layer, making the dimension selection more flexible. 
\citet{guo2020autodis} focused on the embeddings of numerical features. They proposed AutoDis, which automatically discretizes features in numerical fields and maps the resulting categorical features into embeddings.

As for feature interaction, \citet{luo2019autocross} proposed AutoCross that produces high-order cross features by performing beam search in a tree-structure feature space. \citet{song2020towards,khawar2020autofeature,liu2020autofis,xue2020autohash} proposed to automatically discover feature interaction architectures for click-through rate (CTR) prediction. \citet{tsang2020feature} proposed a method to interpret the feature interactions from a source recommendation model and apply them in a target recommendation model.

To the best of our knowledge, we are the first to investigate the automated loss function search for deep recommendation systems.

%% file: 6Conclusion.tex
\vspace{-2.1mm}
\section{Conclusion}
\label{sec:conclusion}
We propose a novel end-to-end framework, AutoLoss, to enhance recommendation performance and deep recommender systems' training efficiency by selecting appropriate loss functions in a data-driven manner. AutoLoss can automatically select the proper loss function for each data example according to their varied convergence behaviors. To be specific, we first develop a novel controller network, which generates continuous loss weights based on the ground truth labels and the DRS' predictions. Then, we introduce a Gumbel-softmax operation to simulate the hard selection over candidate loss functions, which filters out the noisy gradients from suboptimal candidates. Finally, we can select the optimal candidate according to the output from Gumbel-softmax. We conduct extensive experiments to validate the effectiveness of AutoLoss on two widely used benchmark datasets. The results show that our framework can improve recommendation performance and training efficiency with excellent transferability.